\documentclass[12pt]{article}

\usepackage{scicite}

\usepackage{times}

\usepackage[utf8]{inputenc}
\usepackage[T1]{fontenc}
\usepackage{lmodern} 
\usepackage{url}
\usepackage{setspace}

\newenvironment{sciabstract}{%
\begin{quote} \bf}
{\end{quote}}

\newcounter{lastnote}
\newenvironment{scilastnote}{%
\setcounter{lastnote}{\value{enumiv}}%
\addtocounter{lastnote}{+1}%
\begin{list}%
{\arabic{lastnote}.}
{\setlength{\leftmargin}{.22in}}
{\setlength{\labelsep}{.5em}}}
{\end{list}}

\title{D4D-Senegal: The Second Mobile Phone Data for Development Challenge}

\author
{Yves-Alexandre de Montjoye$^{1}$, Zbigniew Smoreda$^{2}$, \\
Romain Trinquart$^{2}$, Cezary Ziemlicki$^{2}$, Vincent D. Blondel$^{3}$\\
\\
\normalsize{$^{1}$Media Lab, Massachusetts Institute of Technology, Cambridge, MA}\\
\normalsize{$^{2}$Orange Labs, France}\\
\normalsize{$^{3}$Université catholique de Louvain, Belgium}\\
\\
}

\begin{document} 


\baselineskip24pt
\onehalfspacing


\maketitle


\begin{sciabstract}
The D4D-Senegal challenge is an open innovation data challenge on anonymous call patterns of Orange's mobile phone users in Senegal. The goal of the challenge is to help address society development questions in novel ways by contributing to the socio-economic development and well-being of the Senegalese population. Participants to the challenge are given access to three mobile phone datasets. This paper describes the three datasets. The datasets are based on Call Detail Records (CDR) of phone calls and text exchanges between more than 9 million of Orange's customers in Senegal between January 1, 2013 to December 31, 2013. The datasets are: (1) antenna-to-antenna traffic for 1666 antennas on an hourly basis, (2) fine-grained mobility data on a rolling 2-week basis for a year with bandicoot behavioral indicators at individual level for about 300,000 randomly sampled users, (3) one year of coarse-grained mobility data at arrondissement level with bandicoot behavioral indicators at individual level for about 150,000 randomly sampled users 
\end{sciabstract}


\section*{Introduction}

There are Big Hopes associated with Big Data: it has been dubbed the oil of the digital economy~\cite{pentland2011society}, the next big thing in medical care~\cite{steinbrook2008personally}, and a vital tool for building smart cities~\cite{cities}. In science, the availability of large-scale behavioral datasets has even been compared to the invention of the microscope~\cite{lazer2009life}. 

There is little doubt that impressive work has already been done by the computational social science and mobile phone research communities. Metadata has, for example, been used to better understand the propagation of malaria, to monitor poverty~\cite{wesolowski2012quantifying, eagle2010network}, to analyse human mobility~\cite{gonzalez2008understanding}, and to study the structure of social communities at a national level~\cite{sobolevsky2013delineating}. Big Data has, however, to be made more broadly available to further realize its promises. Understanding context remains critical, particularly for a sound interpretation and solution of practical questions. Development economists, urban planners, sociologists, and NGOs need to become familiar with this data. ``Inanimate data can never speak for themselves, and we always bring to bear some conceptual framework, either intuitive and ill-formed, or tightly and formally structured, to the task of investigation, analysis, and interpretation''~\cite{gould1981letting}. 

This is why, in 2012, Orange launched the Data For Development challenge in partnership with the University of Louvain and MIT. D4D-Cote d'Ivoire made five months of mobile phone metadata available~\cite{blondel2012data}. The results were impressive: 260 applications from around the world were submitted to access the data and, after three months, more than 80 research papers had been produced~\cite{d4dbook}. These papers covered topics as diverse as optimizing bus routes, analyzing social divisions~\cite{berlingerio2013allaboard}, and studying disease containment policies~\cite{lima2013exploiting}. 

We are now launching, in collaboration with Sonatel Senegal, the second challenge: \textbf{D4D-Senegal}\cite{d4d-senegal} where selected teams will have access to one year of metadata for up to 300,000 people across Senegal. This paper describes the data pre-processing and the three datasets that will be made available, as well as a set of research questions that have been suggested by local partner organizations. More details and the application to participate in the challenge are available at \url{http://www.d4d.orange.com}.

\section*{Data preprocessing}
The Call Detail Records (CDR) have been collected for a year, from January 1 to December 31, 2013. The customer identifiers were anonymized by Sonatel before the data was transferred to Orange Labs who did the preprocessing.

The original dataset contained more than 9 million unique aliased mobile phone numbers. When preparing datasets, we retained only users meeting both of these criteria:
\begin{enumerate}
\item users having more than 75\% days with interactions per given period (biweekly for the second dataset, yearly for the third dataset)
\item users having had an average of less than $1000$ interactions per week. The users with more than $1000$ interactions per week were presumed to be machines or shared phones.
\end{enumerate}

For commercial and privacy reasons, we do not release the real geographical coordinates of the site where BTSs, the mobile network antennas, are located. Note that several BTS can be co-located. We assigned a new position to each site uniformly in its Voronoi cell (the region consisting of all points closer to that antenna than to any other) to make it harder to re-identify users~\cite{ptools}. The \verb|SITE_ARR_LATLON.csv| file contains the new, noisy, latitude and longitude of the site. 

For example:
\begin{verbatim}
site_id,arr_id,lon,lat
1,2,-17.5251,14.74683
2,2,-17.5244,14.74743
3,2,-17.5226,14.7452
4,2,-17.5164,14.74673
\end{verbatim}

\section*{Datasets}
Simply anonymized mobile phone datasets have been shown to be re-identifiable. For instance, it is possible to find a user in a large-scale mobility data using only four spatio-temporal points and coarsening the data only makes it slightly harder~\cite{de2013unique}.

To balance the potential of the data being broadly used with the risks of re-identification we provide three sampled and aggregated datasets for this challenge:
\begin{itemize}
\item \textbf{Dataset 1:} One year of site-to-site traffic for 1666 sites on an hourly basis,
\item \textbf{Dataset 2:} Fine-grained mobility data (site level) on a rolling 2-week basis with bandicoot behavioral indicators at individual level for about 300,000 randomly sampled users meeting the two criteria mentioned before for each 2 week period,
\item \textbf{Dataset 3:} One year of coarse-grained (123 arrondissement level) mobility data with bandicoot behavioral indicators at individual level for about 150,000 randomly sampled users meeting the two criteria mentioned before for a year,
\end{itemize}

Each dataset has been designed to balance utility with privacy, utility beeing the research that can be done with the data while privacy is the potential risk of re-identification of users. Datasets are thus either precise spatially and temporally but limited in the time they span (dataset 2), or aggregated geographically (dataset 3) or across users (dataset 1) but covering a longer period of time. Finally, precomputed indicators are provided to help inform behavioral research. Columns that might help re-identification have been 3-anonymized when binned to remove outliers~\cite{sweeney2002achieving}.

Note that a fourth dataset of synthetic data will be made available in September and will be described in a future paper.

\subsection*{Individual indicators}
Mobility datasets 2 and 3 are supplemented with behavioral indicators from \cite{de2013predicting} computed from metadata using the bandicoot toolbox~\cite{bandicoot}. 

The indicators we provide are:
\begin{itemize}
\item \verb|active_days_callandtext_mean|
\item \verb|active_days_callandtext_sem|
\item \verb|duration_of_calls_mean_mean|
\item \verb|duration_of_calls_mean_sem|
\item \verb|entropy_of_contacts_call_mean|
\item \verb|entropy_of_contacts_call_sem|
\item \verb|entropy_of_contacts_text_mean|
\item \verb|entropy_of_contacts_text_sem|
\item \verb|entropy_of_contacts_callandtext_mean|
\item \verb|entropy_of_contacts_callandtext_sem|
\item \verb|entropy_places_callandtext_mean|
\item \verb|entropy_places_callandtext_sem|
\item \verb|interactions_per_contact_callandtext_mean_mean|
\item \verb|interactions_per_contact_callandtext_mean_sem|
\item \verb|interactions_per_contact_call_mean_mean|
\item \verb|interactions_per_contact_call_mean_sem|
\item \verb|interevents_callandtext_mean_mean|
\item \verb|interevents_callandtext_mean_sem|
\item \verb|interevents_call_mean_mean|
\item \verb|interevents_call_mean_sem|
\item \verb|interevents_text_mean_mean|
\item \verb|interevents_text_mean_sem|
\end{itemize}

Places are in this case sites and nocturnal is defined as 7pm to 7am. A full description of the indicators can be found on the bandicoot document in the data repository and the indicator files have been 3-anonymized on binned data on specific columns to remove outliers~\cite{sweeney2002achieving}.

\subsection*{Dataset 1: Antenna-to-antenna traffic}
This dataset contains the traffic between each site for a year. 

The files \verb|SET1V_M01.csv| through \verb|SET1V_M12.csv| contain monthly voice traffic between sites and are structured as follow:

\begin{itemize}
\item \textbf{timestamp:} day and hour considered in format YYYY-MM-DD HH (24 hours format)
\item \textbf{outgoing\_site\_id:} id of site the call originated from
\item \textbf{incoming\_site\_id:} id of site receiving the call
\item \textbf{number\_of\_calls:} the total number of calls between these two sites during this hour
\item \textbf{total\_call\_duration:} the total duration of all calls between these two sites during this hour
\end{itemize}

For example:
\begin{verbatim}
timestamp, outgoing_site_id, incoming_site_id,...
...number_of_calls, total_call_duration
2013-04-01 00,2,2,7,138
2013-04-01 00,2,3,4,136
2013-04-01 00,2,4,7,121
2013-04-01 00,2,5,13,272
2013-04-30 23,1651,1632,1,3601
2013-04-30 23,1653,575,1,20
2013-04-30 23,1653,1653,2,385
2013-04-30 23,1659,608,1,3601
\end{verbatim}

The files \verb|SET1S_M01.csv| through \verb|SET1S_M12.csv| contain monthly text traffic between sites and are structured as follow:
\begin{itemize}
\item \textbf{timestamp:} day and hour considered in format YYYY-MM-DD HH (24 hours format)
\item \textbf{outgoing\_site\_id:} id of site the text originated from
\item \textbf{incoming\_site\_id:} id of site receiving the text
\item \textbf{number\_of\_sms:} the total number of texts between these two sites during this hour
\end{itemize}

For example:
\begin{verbatim}
timestamp, outgoing_site_id, incoming_site_id, number_of_sms
2013-05-01 00,2,12,6
2013-05-01 00,2,14,1
2013-05-01 00,2,21,1
2013-05-01 00,2,28,9
2013-05-31 23,1653,190,2
2013-05-31 23,1653,314,3
2013-05-31 23,1653,367,8
2013-05-31 23,1653,520,1
2013-05-31 23,1653,558,2
\end{verbatim}

Note that calls spanning multiple time slots are considered to be in the time slot they started in and only calls or texts between Sonatel customers are taken into account.

The latitude and longitude of the sites is provided in \verb|SITE_ARR_LATLON.csv|. 

\subsection*{Dataset 2: Fine-grained mobility}
This second dataset contains the trajectories at site level of about 300,000 randomly selected users meeting the two criteria mentioned before over two-week periods. The site locations are provided in \verb|SITE_ARR_LATLON.csv|.

The files \verb|SET2_P01.csv| through \verb|SET2_P25.csv| contain the \verb|user_id|, \verb|timestamp|, and \verb|site_id| for each of the 25 two-week periods. The second digits of the minutes and all the seconds of the timestamps have been replaced with zeros (format YYYY-MM-DD HH:M0:00)
For each period, a new sample of about 300,000 users was selected and their \verb|user_id| scrambled. Note that this mean that even if a user were to appear in two periods, he would have a different id, and vice versa, the same id in two periods does not mean that it is the same person.

For example:
\begin{verbatim}
user_id,timestamp,site_id 
1,2013-03-18 21:30:00,716
1,2013-03-18 21:40:00,718
1,2013-03-19 20:40:00,716
1,2013-03-19 20:40:00,716
1,2013-03-19 20:40:00,716
1,2013-03-19 20:40:00,716
1,2013-03-19 21:00:00,716
1,2013-03-19 21:30:00,718
1,2013-03-20 09:10:00,705
1,2013-03-21 13:00:00,705
\end{verbatim}

The indicators are computed, for every user, over the course of the two week, and are available in the files \verb|INDICATORS_SET2_P01.csv| through \verb|INDICATORS_SET2_P25.csv|.

\subsection*{Dataset 3: Coarse-grained mobility}
This third dataset contains the trajectories at arrondissement level of 146,352 randomly selected users meeting the two criteria mentioned before on a yearly basis.

\begin{verbatim}
user_id,timestamp,arrondissement_id 
37509,2013-01-29 15:00:00,3
84009,2013-01-14 07:00:00,3
84009,2013-01-14 07:00:00,3
84009,2013-01-14 07:00:00,3
80150,2013-01-27 16:50:00,3
52339,2013-01-09 19:50:00,48
52339,2013-01-06 17:50:00,48
52339,2013-01-13 15:40:00,48
52339,2013-01-03 19:00:00,48
52339,2013-01-07 01:30:00,48
\end{verbatim}

The files \verb|SET3_M01.csv| through \verb|SET3_M12.csv| contain the \verb|user_id|, \verb|timestamp|, and \verb|arrondissement_id| month by month. The second digits of the minutes and all the seconds of the timestamps have been replaced with zeros (format YYYY-MM-DD HH:M0:00)
The indicators are computed, for every user, on a monthly basis. They are available in the files \verb|INDICATORS_SET3_M01.csv| through \verb|INDICATORS_SET3_M12.csv|.

The arrondissement shapefile is provided (\verb|SHAPEFILE_SENEGAL.zip|) as well as a summary table (\verb|SENEGAL_ARR.csv|). 

The summary table contains:
\begin{itemize}
\item \textbf{ARR\_ID:} the arrondissement\_id
\item \textbf{REG:} the name of the region
\item \textbf{DEPT:} the name of the department
\item \textbf{ARR:} the name of the arrondissement
\end{itemize}

For example:
\begin{verbatim}
ARR_ID,REG,DEPT,ARR
1,DAKAR,DAKAR,PARCELLES ASSAINIES
2,DAKAR,DAKAR,ALMADIES
3,DAKAR,DAKAR,GRAND DAKAR
4,DAKAR,DAKAR,DAKAR PLATEAU
5,DAKAR,GUEDIAWAYE,GUEDIAWAYE
6,DAKAR,PIKINE,PIKINE DAGOUDANE
\end{verbatim}

\subsection*{Contextual data}
\begin{itemize}
\item GIS shapefiles for Senegal: Administrative divisions of Senegal shapefiles provided by the ADSN are included in the data package \verb|SHAPEFILE_SENEGAL.zip|
\item Weather data: \url{http://www.wunderground.com/weather-forecast/Senegal.html}
\item Demographic and socio-economic data: \url{http://donnees.ansd.sn/en/BulkDownload}
\item Import/Export data: \url{http://atlas.media.mit.edu/explore/tree_map/hs/export/sen/all/show/2010/}
\item More references at: \url{http://www.d4d.orange.com/en/partners-resources/resources}
\end{itemize}

\section*{Research collaboration}
We strongly encourage developing a scientific collaboration between challenge participants and local teams.  In both its production and interpretation, data is always the result of contingent and contested social practices, the knowledge of national political, cultural and socio-economic context is essential to ask sound research questions and to develop valid results interpretations. 
In order to facilitate this collaboration, the D4D team provide you with a collaborative space on the Sparkboard platform \url{http://d4d.sparkboard.com}, feel free to announce your project and to specify what kind of competencies and collaboration you would be interested in.


\bibliography{scibib}

\bibliographystyle{Science}


\begin{scilastnote}
\item The authors would like to thanks Kevin Mustelier and Luc Rocher for their help with the Bandicoot toolbox

\end{scilastnote}


\clearpage

\end{document}